\documentclass[sigconf]{acmart}

\AtBeginDocument{%
  }

\usepackage{listings}
\lstdefinestyle{shell}{
  basicstyle=\ttfamily\scriptsize,
  breaklines=true,
  frame=single,
  backgroundcolor=\color{black!4},
  commentstyle=\color{black!50},
  morecomment=[l]{\#},
  belowskip=0pt,
}

\definecolor{rediBlue}{HTML}{2563EB}
\definecolor{setgoGreen}{HTML}{059669}
\definecolor{setgoGreenLight}{HTML}{D1FAE5}
\definecolor{aidrinAmber}{HTML}{D97706}
\definecolor{rawRed}{HTML}{DC2626}
\definecolor{textDark}{HTML}{111827}
\definecolor{textMid}{HTML}{6B7280}
\definecolor{arcGray}{HTML}{9CA3AF}

\copyrightyear{2026}
\acmYear{2026}
\setcopyright{usgov}
\acmConference[SSDBM 2026]{The International Conference on Scalable Scientific Data Management 2026}{August 11--13, 2026}{San Diego, CA, USA}
\acmBooktitle{The International Conference on Scalable Scientific Data Management 2026 (SSDBM 2026), August 11--13, 2026, San Diego, CA, USA}
\acmDOI{10.1145/3828820.3828827}
\acmISBN{979-8-4007-2708-5/26/08}

\begin{document}

\title{SetGo: Metadata Readiness for Scientific AI Datasets}

\author{Sean R. Wilkinson}
\orcid{0000-0002-1443-7479}
\affiliation{%
  \institution{Oak Ridge National Laboratory}
  \city{Oak Ridge}
  \state{Tennessee}
  \country{USA}
}
\email{wilkinsonsr@ornl.gov}

\author{Polina Shpilker}
\orcid{0000-0002-6761-7326}
\affiliation{%
  \institution{Oak Ridge National Laboratory}
  \city{Oak Ridge}
  \state{Tennessee}
  \country{USA}
}
\email{shpilkerph@ornl.gov}

\author{Wesley Brewer}
\orcid{0000-0002-3639-3956}
\affiliation{%
  \institution{Oak Ridge National Laboratory}
  \city{Oak Ridge}
  \state{Tennessee}
  \country{USA}
}
\email{brewerwh@ornl.gov}

\renewcommand{\shortauthors}{Wilkinson et al.}

\begin{abstract}
Scientific datasets intended for AI use require both computational readiness
for model training and metadata readiness for discovery, sharing, and reuse.
The Readiness Engine for Data Integration (REDI) addresses computational
readiness, but no corresponding tool evaluates whether a dataset's metadata
are sufficiently complete, governed, and standards-compliant for publication
and agent-based consumption. Existing FAIR assessors operate only on published
repository records, and no single system covers FAIR compliance, licensing,
provenance, governance, reproducibility, and catalog readiness together.

We present \textbf{SetGo}, an open-source Python toolkit
that assesses and repairs metadata readiness across these six dimensions before
a dataset is published or archived. Applied to four scientific corpora, SetGo
surfaces deficiencies that general-purpose tools do not detect: ERA5 climate
metadata scores 4\% on ACDD~1.3 compliance; materials datasets fail OPTIMADE
species-definition requirements; and PDB-derived proteomics data carries
licensing terms incompatible with standard SPDX identifiers. Guided enrichment
raises overall FAIR scores from 52--57\% to 81--91\%, and a single
\texttt{setgo publish} command pushes to Hugging Face Hub, CKAN, or
OpenMetadata with ML~Commons Croissant~1.0 metadata sidecars.

To support interactive and automated workflows, SetGo integrates with coding agents powered by large language models (LLMs) through a \texttt{/setgo} skill that enables natural-language execution of the full assess--enrich--publish loop, with user involvement limited to supplying missing metadata values.
 \end{abstract}



\maketitle

\section{Introduction}
Artificial intelligence is increasingly applied across scientific domains,
including climate modeling, protein structure prediction, materials discovery,
and plasma physics simulation \cite{nguyen2023climax, ahdritz2024openfold,
pasini2024scalable}. These efforts depend on datasets that are not only
computationally prepared for model training but also accompanied by metadata
rich enough for discovery, governance, reuse, and agent-based consumption. The
FAIR principles \cite{wilkinson2016fair} provide a widely adopted stewardship
framework, including across U.S.~Department of Energy (DOE) user facilities.
Yet FAIR compliance alone does not ensure that datasets are ready for
downstream AI workflows. In practice, data may satisfy FAIR criteria while
lacking the provenance, licensing clarity, or catalog registration needed for
responsible reuse; conversely, computationally prepared datasets often omit
these metadata elements entirely \cite{wilkinson2022fworkflows}.

The Data Readiness for AI principles (DRAI) \cite{brewer2026datareadiness} formalize this distinction: \emph{computational
readiness} covers the transformations needed for model training, while
\emph{metadata readiness} concerns the governance, provenance, and discoverability required for compliant reuse. The Readiness Engine for Data
Integration (REDI) implements a unified five-stage pipeline for computational
readiness \cite{redi2026}. However, existing tools provide only partial support
for evaluating metadata readiness prior to publication, and to our knowledge no
single system assesses the full combination of FAIR sub-principles, licensing,
provenance, governance policies, reproducibility, and multi-catalog publication
workflows required for scientific datasets.

We introduce \textbf{SetGo}, an open-source Python toolkit (v0.1.0, Apache-2.0)%
\footnote{Source code: \url{https://code.ornl.gov/drai/setgo}, DOI: \href{https://doi.org/10.11578/dc.20260708.9}{10.11578/dc.20260708.9}}
that fills this gap by assessing and improving metadata readiness before
datasets are published or archived. A single
\texttt{setgo publish} command pushes enriched datasets to Hugging Face Hub,
CKAN systems, or OpenMetadata installations and produces ML~Commons
Croissant~1.0 metadata sidecars. SetGo integrates into automated high
performance computing (HPC) data pipelines and continuous integration and
deployment (CI/CD) workflows, and its domain-aware validators support climate
science (CF Conventions, ACDD~1.3), materials science (CIF, OPTIMADE), life
sciences (MIAME, BioSchemas), and nuclear fusion (IMAS). Although developed as
a companion to REDI, SetGo accepts any \texttt{metadata.json} record conforming
to its documented, open schema and integrates into pipelines beyond REDI.

\begin{figure*}[t]
  \centering
  \includegraphics[width=0.72\textwidth]{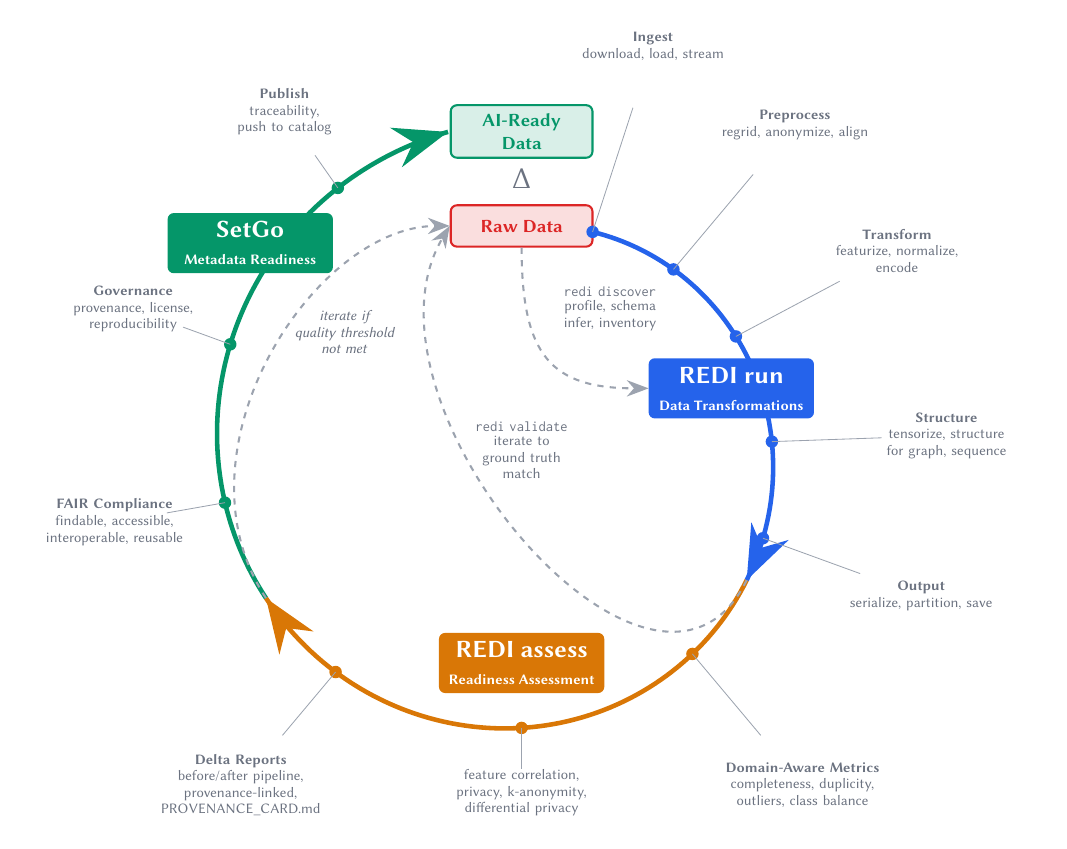}
  \caption{The REDI--SetGo iterative data readiness lifecycle \cite{redi2026}.
    REDI's five-stage pipeline prepares raw data for model training; SetGo makes
    the output publication-ready for governed, discoverable, and agent-based
    consumption ($\Delta$ denotes readiness improvement). Datasets below the
    readiness threshold are returned to REDI for further refinement.}
  \label{fig:lifecycle}
\end{figure*}

Figure~\ref{fig:lifecycle} shows SetGo within an iterative data-readiness
lifecycle alongside REDI~\cite{redi2026}, jointly addressing computational
preparation and metadata enrichment. We evaluate SetGo's ability to identify,
repair, and validate metadata deficiencies prior to publication, rather than
claiming to establish absolute metadata quality.

The contributions of this work are:
\vspace{-4pt}
\begin{itemize}
\item A unified six-dimensional metadata-readiness framework with domain-aware
      FAIR assessment across all 15 sub-principles, quantitative scoring, and
      letter-grade thresholds;
\item Integrated validation of SPDX (Software Package Data Exchange) licenses, W3C~PROV-O provenance, and
      environment reproducibility, with automated multi-catalog publication and
      Croissant~1.0 metadata sidecar generation;
\item A skill-file interface that enables coding agents using large language models (LLMs) to execute the full assess–enrich–publish workflow through natural-language interaction; and
\item An open-source implementation evaluated across four scientific domains.
\end{itemize}
\vspace{-4pt}
 
\vspace{-4pt}
\section{Related Work}
\subsection{Data Readiness Frameworks}
Lawrence~\cite{Lawrence2017} introduced readiness bands describing how raw
records become suitable for a specific task; DRAI~\cite{brewer2026datareadiness}
extended this with a five-stage Ingest-Preprocess-Transform-Structure-Output (IPTSO) pipeline and levels from RAW to
FULLY\_AI\_READY. Later surveys note that many readiness approaches focus on
computational or structural transformations and provide limited support for
determining whether metadata are complete, governed, or suitable for
publication~\cite{hiniduma2025data}. Domain-targeted tools such as AIDRIN~\cite{hiniduma2024ai}
and Bridge2AI~\cite{clark2024} assess data quality and governance in specific
domains, but neither provides general, pre-publication metadata readiness
assessment across scientific datasets.

\vspace{-4pt}
\subsection{FAIR and Metadata-Assessment Tools}
FAIR compliance tools such as F-UJI~\cite{devaraju2021automated} and the FAIR
Evaluator analyze published datasets by querying registered repository records.
These systems provide valuable feedback once metadata are already online, but
do not assist with pre-publication diagnosis or repair, integrate with local
dataset directories, or surface governance and licensing gaps. FAIR has also
been extended to computational workflows~\cite{wilkinson2025fairworkflows},
reinforcing the need for readiness checks beyond published datasets. SetGo
operates earlier: it evaluates local metadata, recommends targeted
improvements, and connects those improvements directly to catalog export.

\vspace{-4pt}
\subsection{Provenance and Experiment Tracking}
Provenance and experiment-tracking systems address important but isolated
dimensions of metadata readiness. MLflow~\cite{zaharia2018mlflow} captures model
parameters and artifacts but does not record dataset-level lineage or enforce
governance constraints. Flowcept~\cite{flowcept} provides fine-grained,
workflow-level provenance for REDI but does not assess whether the resulting
metadata meet publication standards. DataLad~\cite{halchenko2021datalad} offers version-controlled management of
data and code but does not evaluate FAIR compliance, licensing completeness,
or catalog readiness.

\vspace{-4pt}
\subsection{Catalog Standards and Data Quality}
The ML~Commons Croissant~1.0 specification defines a JSON-LD schema for
describing ML datasets and promotes consistent structure across platforms such as
Hugging Face, Kaggle, and OpenML. Although Croissant standardizes key metadata
fields, it does not diagnose missing information or guide users through preparing
datasets for compliant publication. Data-quality frameworks like Great Expectations identify anomalies and validate
pipeline behavior but do not assess FAIR compatibility, validate SPDX licensing,
or check governance requirements.

\vspace{-4pt}
\subsection{Agentic and LLM-Assisted Tools}
LLM-based coding agents streamline interaction with command-line tools by
executing natural-language instructions. Systems such as Claude
Code~\cite{claudecode2025} rely on structured skill files to expose tool
interfaces reproducibly, but do not assess metadata completeness or governance
constraints on their own. SetGo’s \texttt{SKILL.md} file and \texttt{/setgo}
command integrate agent support directly into the readiness workflow, keeping
all checks deterministic and auditable.

\vspace{-4pt}
\subsection{Synthesis}
To our knowledge, no prior system evaluates FAIR sub-principles, governance
policies, licensing, provenance completeness, reproducibility, and
catalog-ready fields within a single pre-publication workflow. SetGo fills this
gap by coupling assessment with guided enrichment and automated catalog export.
 
\vspace{-4pt}
\section{SetGo Design}
\label{sec:design}
SetGo is organized as six independent  assessment modules, a
catalog-integration layer, and a reporting subsystem. The toolkit accepts a
dataset directory as its primary input, parsing a local \texttt{metadata.json}
file (the standard REDI output format) along with any co-located data files.

The CLI exposes commands for individual checks (\texttt{setgo fair},
\texttt{setgo check}) and for full-lifecycle execution. The
\texttt{setgo assess} command runs all modules and compiles results into a
single readiness report, while \texttt{setgo run} extends this by publishing
enriched metadata to the selected catalog target.

A \texttt{ReadinessReport} aggregates the four scored dimensions into a
composite score on a 0--1 scale with equal weighting; provenance and
traceability are recorded structurally rather than scored. Letter-grade
thresholds (A $\geq 90\%$, B $\geq 80\%$, C $\geq 70\%$, D $\geq 60\%$,
F $<60\%$), following conventional grading conventions, provide a concise and
actionable summary of metadata readiness. Datasets that reach grade~C are
flagged as AI-ready (\texttt{is\_ready = True}). Reports can be exported as
JSON, Markdown, or styled HTML, and in CI/CD settings \texttt{setgo check}
returns exit code~0 once a dataset meets the configured threshold.

\vspace{-4pt}
\subsection{FAIR Assessment}
\label{sec:fair}
SetGo evaluates all 15 FAIR sub-principles across the four FAIR categories. Each
principle is scored on a four-point scale: 0 (not evaluated), 1 (non-compliant),
2 (partially compliant), or 3 (fully compliant). Category and overall scores are
computed as the mean of the \emph{assessed} sub-principle scores; a sub-principle
that cannot be evaluated (score 0) is omitted from the mean, not counted as a
zero.

A set of detection heuristics supports common metadata patterns. For F1, SetGo
searches for DOI, Archival Resource Key (ARK), Handle, and UUID patterns; for I1, file extensions are
mapped to interoperability levels, with higher scores assigned to formats such
as \texttt{.nc}, \texttt{.hdf5}, and \texttt{.parquet}. Licensing checks
recognize SPDX identifiers (R1.1), and R1.2 inspects fields documenting
provenance, authorship, and source URLs. Domain-aware validators extend these
checks: CF Conventions and ACDD (Attribute Convention for Data Discovery)~1.3
for climate, CIF and OPTIMADE~v1.1 for materials,
Protein Data Bank (PDB)/FASTA validators aligned with MIAME/MINSEQE for life sciences,
and IMAS for nuclear fusion. These validators are pluggable, and new ones can be
registered through a YAML configuration file.

Assessment produces a \texttt{FAIRRecommender} output that ranks potential
improvements by their expected impact on the composite FAIR score.

\vspace{-4pt}
\subsection{Compliance and Licensing}
\label{sec:compliance}
The governance module evaluates datasets against configurable data policies,
each represented as a collection of typed rules. Three built-in levels are
provided: MINIMAL (basic metadata presence), STANDARD (license, provenance, and
data-quality requirements), and STRICT (FAIR score $\geq 0.7$ plus complete
provenance and a published license). Custom policies can be defined in YAML.
Each rule yields a structured verdict (COMPLIANT, PARTIAL, or NON\_COMPLIANT)
with a human-readable message gathered into a \texttt{ComplianceReport}.

Licenses are validated against the SPDX registry, which covers more than twenty
widely used data licenses (including CC0-1.0, CC-BY-4.0, and ODbL-1.0). SetGo
propagates the validated license conditions into Hugging Face dataset cards and
Croissant~1.0 metadata fields to maintain consistency across catalog targets.

\vspace{-4pt}
\subsection{Provenance, Reproducibility, and Traceability}
\label{sec:provenance}
Provenance is recorded using a W3C PROV-O compatible model comprising entities
(data files with MIME types and SHA-256 checksums), activities (processing steps
with declared inputs and outputs), and agents (persons or systems). The lineage
tracker captures each transformation and exports the resulting graph as
PROV-JSON, PROV-JSON-LD, or PROV-N, which can be embedded directly into catalog
records upon publication.

SetGo's provenance model complements Flowcept's workflow-level instrumentation.
Where Flowcept tracks fine-grained state transitions inside REDI's transformation
pipeline, SetGo constructs a dataset-level provenance document suitable for
catalog metadata. The reproducibility module captures the software environment
at assessment time (Python version, OS, installed packages, hardware
identifiers) and verifies artifact integrity using SHA-256 checksums.

Traceability links datasets to AI models through typed relationships (training,
validation, evaluation). Integration with MLflow enables bidirectional traversal
between model outputs and training-data provenance.

\vspace{-4pt}
\subsection{Catalog Integration}
\label{sec:catalogs}
The \texttt{setgo publish} command pushes datasets to catalog systems through a
pluggable registry. Exporters are loaded lazily to avoid unnecessary
dependencies. A YAML configuration passed to \texttt{setgo run} specifies both
the assessment policy and the intended publication target, enabling automated
deployment in CI/CD settings.

Supported targets include Hugging Face Hub (with dataset cards, Parquet
conversion, and optional access controls), CKAN (public or private instances via
REST), and OpenMetadata (including DOE-operated deployments such as the AmSC
Data Catalog). SetGo also generates Croissant~1.0 metadata sidecars (JSON-LD
with checksums and field-level annotations), which serve as portable complements
to catalog-native formats. License conditions propagate uniformly across all
targets: attribution notices appear in dataset cards, SPDX URLs populate
JSON-LD fields, and non-commercial restrictions are flagged in catalog records.

\vspace{-4pt}
\subsection{Agentic Interface}
\label{sec:agentic}
SetGo supports natural-language interaction via LLM-based coding agents,
allowing domain scientists to run the full assess--enrich--publish workflow
without invoking CLI commands directly. The toolkit includes a
\texttt{SKILL.md} file that describes commands, output schemas, and
recommended workflows for use by coding agents such as Claude Code. Executed within such an agent, a
\texttt{/setgo} slash command provides a direct entry point for agent-driven
execution. SetGo requires explicit user confirmation before adding
authoritative metadata such as persistent identifiers, licenses, or
provenance sources, ensuring that these fields are never autonomously
fabricated.

An agent calls \texttt{setgo fair {-}{-}style json}, identifies gaps, patches
\texttt{metadata.json}, and repeats until the threshold is met; natural-language
orchestration lets agents prioritize gaps and phrase prompts appropriately for
the scientist while validators keep all checks deterministic and auditable. This workflow supports DOE
Genesis Mission~\cite{genesismission} data card production;
Figure~\ref{fig:cli} illustrates the enrichment loop on an ERA5 climate
fixture, showing how minimal user input enables full metadata preparation
and publication.

\begin{figure*}[t]
\begin{minipage}[t]{0.47\textwidth}
\begin{lstlisting}[style=shell]
$ setgo fair ./climax-test -d climate
  Domain:   climate
  Overall:  57%

  FINDABLE       42%
  ACCESSIBLE     78%
  INTEROPERABLE  56%
  REUSABLE       58%

  -> Assign a persistent identifier (DOI)
  -> Specify an SPDX license identifier
  -> Document data provenance
\end{lstlisting}
\end{minipage}%
\hfill%
\begin{minipage}[t]{0.47\textwidth}
\begin{lstlisting}[style=shell]
User:  Check FAIR readiness and publish this dataset.
Agent: Score 57% -- not AI-ready. Three gaps:
       F1 (no DOI), R1.1 (no license), R1.2 (no provenance).
       Persistent identifier?
User:  10.5281/zenodo.XXXXXXX
Agent: License?
User:  CC-BY-4.0
Agent: Provenance source?
User:  ERA5 reanalysis (ECMWF)
Agent: Patched. Re-checking...
       FAIR 91%, governance PASS (5/5).
       Publishing... croissant.json (CC-BY-4.0).
\end{lstlisting}
\end{minipage}
\caption{\textbf{Left:} Raw \texttt{setgo fair} output for an ERA5 climate
dataset, showing a 57\% initial FAIR score and three NOT\_COMPLIANT principles.
\textbf{Right:} An LLM agent guided by \texttt{SKILL.md} executes the full
assess--enrich--publish loop from a single natural-language request, calling
\texttt{setgo fair {-}{-}style json} and \texttt{setgo publish} internally while
prompting the scientist only for the three missing metadata fields (transcript
condensed for space; values taken from an actual run).}
\label{fig:cli}
\end{figure*} 
\vspace{-4pt}
\section{Evaluation}
\label{sec:eval}
We evaluate whether SetGo can identify, repair, and validate metadata
deficiencies prior to publication across the four scientific domains used to
benchmark REDI~\cite{redi2026}: climate (ERA5/ClimaX~\cite{nguyen2023climax}),
proteomics (OpenFold~\cite{ahdritz2024openfold}), materials science
(OC22/HydraGNN~\cite{pasini2024scalable}), and nuclear fusion (XGC1). These
corpora span from 4.3~TB (proteomics) to 106~TB (7,089 XGC1 runs) and reflect
the scale and heterogeneity typical of HPC-driven workflows.

Let $s_i \in \{1,2,3\}$ denote the compliance score assigned to an \emph{assessed}
FAIR sub‑principle $i$, and let $A \subseteq \{1,\dots,15\}$ be the set of
sub-principles the assessor can evaluate for a given record ($s_i = 0$ marks a
sub-principle as not evaluated). The normalized FAIR score is the mean over the
assessed set,
$S_{\mathrm{FAIR}} = \frac{1}{|A|}\sum_{i \in A}\frac{s_i}{3}$, with each assessed
sub-principle weighted equally. Sub-principles that cannot be evaluated from a
local, pre-publication record (e.g., A2) are excluded from the mean; they are not
counted as zeros. Equal weighting was selected to avoid over‑prioritizing any
single FAIR dimension and to maintain interpretability across heterogeneous
domains.

To evaluate SetGo's ability to identify and repair metadata deficiencies, we adopt a fixture-based testing methodology. Rather than rerunning the full REDI pipeline on multi-terabyte corpora for each evaluation, we construct paired metadata records representing the state of each dataset before and after guided enrichment. This approach ensures that reported scores are deterministic and independently reproducible without requiring access to restricted institutional storage systems or incurring the computational cost of end-to-end pipeline execution. Both fixture sets are included in the open-source repository so that any reader can reproduce the reported scores by running \texttt{setgo assess} directly.

SetGo evaluates each dataset through its \texttt{metadata.json} record produced
by REDI, with domain validators inspecting co‑located data files when present.
Before-state fixtures are drawn directly from REDI runs, with internal paths and
credentials removed but all readiness‑relevant metadata preserved. After-state
fixtures reflect the identifiers, licenses, and provenance links that a
scientist supplies through SetGo’s guided enrichment workflow, reproducing the
outputs of actual enrichment sessions; both fixture sets are included in the
open-source repository for reproducibility. Score improvements therefore follow
SetGo’s ranked recommendations: the before-state represents common pipeline
outputs in which metadata remain incomplete, and the after-state demonstrates
the ceiling achievable when missing information is supplied. Improvements in the
remaining four dimensions are computed deterministically from the enriched
metadata record. Direct comparison to existing FAIR tools (F-UJI or the FAIR
Evaluator) is not possible because they require a registered repository URL and
do not operate on local, unpublished directories.

SetGo requires explicit user confirmation for authoritative metadata fields such
as persistent identifiers, licensing information, and provenance sources,
preventing autonomous fabrication of publication‑critical metadata. All scoring rules are grounded in external published standards (FAIR
sub-principles, SPDX, W3C~PROV-O, CF, ACDD, OPTIMADE, IMAS) rather than
learned or opaque metrics, so score improvements reflect genuine metadata
additions.

\vspace{-4pt}
\subsection{FAIR Assessment}
Table~\ref{tab:fair} reports FAIR category scores before and after applying
SetGo’s recommendations. Initial overall FAIR scores range from 52\% (Fusion) to
57\% (Climate and Proteomics), reflecting datasets that contain basic structural
metadata but lack persistent identifiers, machine-readable licenses, or
provenance information. Findable scores are uniformly lowest at 42\%, due to the
absence of DOIs or ARKs. Accessible scores cluster at 78\% because data remain
reachable via institutional storage systems. Interoperable and Reusable scores
vary by domain; Fusion and Materials score lower on Interoperability because
particle‑simulation and atomistic-structure formats expose fewer fields that map
cleanly to generic FAIR sub-principles. Climate and Proteomics share identical
before-state scores because both fixtures arrive from REDI pipelines with the
same structural gaps (no persistent identifier, no machine‑readable license, and
no provenance record) despite differing domain validators.

After enrichment, FAIR scores improve substantially. The climate dataset reaches
91\%, driven by added identifiers, licensing, and provenance metadata.
Proteomics reaches a lower ceiling (81\%) because PDB‑derived licensing
constraints prevent full catalog registration, leaving F4 (indexed in a
searchable resource) non-compliant.

\begin{table}[t]
 \caption{FAIR category scores (\%) before and after SetGo-guided metadata
 enrichment; scores are percentages of the maximum possible (3 per
 sub-principle). F~=~Findable, A~=~Accessible, I~=~Interoperable,
 R~=~Reusable. Overall is the mean of the assessed sub-principle scores (each
 weighted equally; sub-principles the assessor cannot evaluate, e.g.\ A2, are
 omitted).}
 \label{tab:fair}
 \small
 \begin{tabular}{lcccccc}
 \toprule
 Domain & Phase & F & A & I & R & Overall \\
 \midrule
 Climate    & Before & 42 & 78 & 56 & 58 & 57 \\
            & After  & 83 & 92 & 89 & 100 & 91 \\
 \addlinespace
 Proteomics & Before & 42 & 78 & 56 & 58 & 57 \\
            & After  & 67 & 100 & 78 & 83 & 81 \\
 \addlinespace
 Materials  & Before & 42 & 78 & 44 & 58 & 55 \\
            & After  & 75 & 92 & 78 & 100 & 87 \\
 \addlinespace
 Fusion     & Before & 42 & 78 & 44 & 50 & 52 \\
            & After  & 75 & 92 & 67 & 92 & 82 \\
 \bottomrule
 \end{tabular}
\end{table}

Domain-aware validators surface deficiencies that general-purpose FAIR tools do
not detect. For the materials dataset, the OPTIMADE validator reports a missing
species-definition field required for catalog interoperability. For the ERA5‑
based climate fixture, \texttt{setgo fair -d climate} reveals two issues:
CF‑Conventions compliance at 75\% (missing \texttt{Conventions} and
\texttt{standard\_name} fields) and ACDD~1.3 compliance at 4\% (absent title,
summary, and geospatial attributes). The \texttt{FAIRRecommender} ranks these
gaps by their expected contribution to the composite score.

\vspace{-4pt}
\subsection{Governance Compliance}
Applying the STRICT policy to REDI outputs highlights common failure modes
across all domains: missing license declarations, incomplete provenance records,
and FAIR scores below the 0.70 threshold. After enrichment, all four datasets
meet STANDARD compliance, and three achieve STRICT compliance. The proteomics
fixture remains an exception due to PDB-derived licensing terms, which do not
align with a standard SPDX identifier and require institutional review. This
case illustrates SetGo’s ability to surface real governance constraints rather
than silent failures.

\vspace{-3pt}
\paragraph{Non-FAIR dimensions.}
After enrichment, SPDX license conditions propagate into all catalog targets
and Croissant~1.0 sidecars. SetGo exports a W3C PROV-O provenance document per
fixture and records the assessment environment (Python version, OS, packages,
SHA-256 checksums) for reproducibility. Four typed model-dataset linkages
(ClimaX, OpenFold, HydraGNN, XGC1) enable bidirectional audit trails.

\vspace{-4pt}
\subsection{Agentic Enrichment and Catalog Export}
Fig.~\ref{fig:cli} illustrates the enrichment workflow on the climate fixture. A
single natural‑language request prompts the agent to run \texttt{setgo fair
{-}{-}style json}, identify missing metadata (F1, R1.1, R1.2), and request the
necessary DOI, license, and provenance fields. The agent patches
\texttt{metadata.json}, re-runs assessment, and verifies a 91\% FAIR score
together with a passing governance check.

When a scientist supplies an unrecognized license string, SetGo’s SPDX validator
flags it, and the agent re-prompts with valid options. If a required value
cannot be inferred—such as a DOI not yet minted—the agent marks it as a
prerequisite rather than introducing a placeholder.

\vspace{-3pt}
\paragraph{Catalog export.}
Running \texttt{setgo publish} on an enriched fixture produces a Croissant~1.0
JSON-LD sidecar validated with \texttt{mlcroissant}~1.1.0, including SHA‑256
checksums, SPDX license URLs, field-level annotations, and creator metadata
derived from \texttt{metadata.json}. For Hugging Face Hub, SetGo generates a
dataset card with inferred license notices. The same metadata record supports
export to CKAN and OpenMetadata without additional authoring; a single YAML file
specifies the target, and SetGo handles authentication, record creation, and
file attachment.

\vspace{-3pt}
\paragraph{Assessment performance.}
Because SetGo operates on co-located samples and \texttt{metadata.json} records 
rather than full datasets, assessment latency is independent of raw data volume.
On the four evaluation fixtures, \texttt{setgo assess} completes in well under
100~ms for metadata-only inputs; when a co-located data file is present---as in
the climate fixture, where CF Conventions and ACDD 1.3 validators inspect a
sample \texttt{.nc} file---latency remains under 500~ms. The 106~TB XGC1 corpus
is cited for scientific context; its \texttt{metadata.json} record is the sole
input to assessment, which completes in the same sub-100~ms range.

\vspace{-4pt}
\subsection{Threats to Validity}
SetGo’s readiness and FAIR-aligned scores are derived from heuristic,
rule‑based assessments implemented within the framework itself, so improvements
after enrichment partially reflect optimization against SetGo’s own scoring
criteria. To mitigate this risk, the scoring rules are grounded directly in
published FAIR sub-principles, SPDX validation rules, W3C PROV-O structures, and
domain standards (e.g., CF Conventions, ACDD, OPTIMADE, IMAS) rather than
opaque or learned metrics. External FAIR assessment tools such as F‑UJI and the
FAIR Evaluator could not be used for direct comparison because they require
publicly registered repository records and do not operate on unpublished local
datasets. Consequently, this evaluation emphasizes pre‑publication metadata
readiness rather than cross‑tool benchmarking. Future work will include expert
annotation studies, cross-tool comparison with F-UJI on published records, and
downstream discoverability analyses to further validate the scoring methodology.

\vspace{-3pt}
\paragraph{Limitations.}
SetGo v0.1.0 treats REDI’s \texttt{metadata.json} as its canonical input; other
pipelines require lightweight adapters. The SPDX registry covers common open
licenses but flags jurisdiction-specific or proprietary variants for manual
review. Interactive enrichment requires user-supplied values for essential
fields (title, description, creators, license). The \texttt{SKILL.md} workflow
helps surface these gaps but cannot infer values absent from context. A
prospective user study is needed to confirm that score improvements translate to
measurably better downstream discoverability and reuse; this is deferred to
future work.
 
\vspace{-4pt}
\section{Conclusion}
We presented SetGo, an open-source Python toolkit for assessing and managing the
metadata readiness of scientific AI datasets. By evaluating FAIR compliance,
governance, licensing, provenance, reproducibility, and traceability, and
linking those assessments directly to catalog publication, SetGo connects REDI’s
training-ready outputs to the metadata properties needed for governed,
discoverable, and agent-accessible datasets. The \texttt{/setgo} slash command
and the \texttt{SKILL.md} interface allow LLM-based agents to execute the full
assess--enrich--publish loop with minimal user intervention, giving domain
scientists a practical path to producing FAIR-compliant, catalog-ready metadata
without deep expertise in standards or policy.

Future directions include automated metadata repair (unattended persistent identifier minting,
SPDX registration, and provenance linking), expanded domain and catalog coverage
(neutron science, HPC telemetry, WorkflowHub~\cite{gustafsson2025}), and direct integration of
SetGo’s readiness gate into REDI’s output stage for a unified raw-to-catalog
pipeline.

As agent-access protocols such as WebMCP~\cite{webmcp} mature, the persistent
identifiers, machine-readable licenses, and Croissant~1.0 metadata that SetGo
produces will enable AI agents to locate and consume scientific datasets
directly, advancing the FAIR vision of machine-actionable data. 
\vspace{-4pt}
\begin{acks}
This research used resources of the Oak Ridge Leadership Computing Facility at Oak
Ridge National Laboratory, which is supported by the Office of Science of the
U.S.\ Department of Energy under Contract No.\ DE-AC05-00OR22725.
AI writing assistance (Anthropic Claude) was used for language editing throughout.
\end{acks}

\bibliographystyle{ACM-Reference-Format}

\end{document}